\documentclass{iaus}
\usepackage{graphics}
\input{psfig}
  \checkfont{eurm10}
  \iffontfound
    \IfFileExists{upmath.sty}
      {\typeout{^^JFound AMS Euler Roman fonts on the system,
                   using the 'upmath' package.^^J}%
       \usepackage{upmath}}
      {\typeout{^^JFound AMS Euler Roman fonts on the system, but you
                   dont seem to have the}%
       \typeout{'upmath' package installed. iaus.cls can take advantage
                 of these fonts,^^Jif you use 'upmath' package.^^J}%
      }
  \else
  \fi

  \checkfont{msam10}
  \iffontfound
    \IfFileExists{amssymb.sty}
      {\typeout{^^JFound AMS Symbol fonts on the system, using the
                'amssymb' package.^^J}%
       \usepackage{amssymb}%

      }{}
  \fi

  \IfFileExists{amsbsy.sty}
    {\typeout{^^JFound the 'amsbsy' package on the system, using it.^^J}%
     \usepackage{amsbsy}}
    {}

\newsavebox{\astrutbox}
\sbox{\astrutbox}{\rule[-5pt]{0pt}{20pt}}

\def\HI{H{\,\small I}}
\def\OII{O{\,\small II}}
\newcommand{\ltsima} {$\; \buildrel < \over \sim \;$}
\newcommand{\gtsima} {$\; \buildrel > \over \sim \;$}
\newcommand{\lta} {\lower.5ex\hbox{\ltsima}}
\newcommand{\gta} {\lower.5ex\hbox{\gtsima}}

\title[The Interplay among Black Holes, Stars and ISM in Galactic 
       Nuclei]{Fast Outflows of Neutral Hydrogen in Radio Galaxies}

\author[T.A. Oosterloo {\it et al.\/}]%
{T.A. Oosterloo$^1$, R. Morganti$^1$,
B.H.C. Emonts$^2$\break \and C.N. Tadhunter$^3$}

\affiliation{$^1$Netherlands Foundation for Research in Astronomy, Postbus 2,
NL-7990 AA, Dwingeloo, NL
email: oosterloo@astron.nl\\[\affilskip]
$^2$ Kapteyn Astronomical Institute, RuG, Landleven 12, 9747 AD,
Groningen, NL \\[\affilskip] 
$^3$ Dept.\ Physics \& Astronomy,
University of Sheffield, S7 3RH, UK}

\pubyear{2004}
\volume{222}
\pagerange{1--8}
\date{?? and in revised form ??}
\jname{The Interplay among Black Holes, Stars and ISM \\in Galactic Nuclei}
\editors{Th. Storchi Bergmann, L.C. Ho \& H.R. Schmitt, eds.}
\begin{document}

\maketitle

\begin{abstract}
AGN activity is known to drive fast outflows of gas. We report the discovery of
fast outflows of {\sl neutral} gas with velocities over 1000 km/s in a number
of radio galaxies.  In the best studied object, 3C~293, the kinematical
properties of the neutral and ionised outflows are similar, indicating a common
origin. Moreover, the outflow appears to be located near the radio lobes and
not near the nucleus. This suggests that the interaction between the radio jet
and the ISM is driving the outflow.
\end{abstract}

\firstsection 
\section{Discovery of fast \HI\ outflows}

AGN activity is increasingly recognised to play an important role in the
evolution of galaxies.  Particularly important in this respect are the gas
outflows that can be generated by the nuclear activity and the significant
effect that these outflows can have (together with the outflows generated by
starbursts) on the interstellar medium (ISM) of a galaxy and hence on its
evolution.  It is, therefore, particularly interesting that in a number of
radio galaxies we have now discovered evidence for fast outflows of neutral
gas having velocities larger than 1000 km/s.  This shows that, despite the
highly energetic phenomena that must be associated with these fast outflows,
part of the gas remains, or becomes again, neutral.  The discovery of fast
outflows of neutral hydrogen in radio galaxies was done using \HI\
observations performed with the Westerbork Synthesis Radio Telescope (WSRT),
taking advantage of the new 20 MHz broad-band system and of its excellent
bandpass stability.  The first fast \HI\ outflow was found in the radio galaxy
3C~293 (Fig.\ 1, Morganti et al.\ 2003), but observations carried out recently
(again using the WSRT) have revealed fast \HI\ outflows in five objects (of
the 10 observed) with line widths ranging from 800 up to 2000 km/s.

The optical depths of the broad, shallow absorption are low (typically around
0.001).  Nevertheless, due to the large velocity width, the implied column
densities are high: around $10^{21}$ cm$^{-2}$ for a $T_{\rm spin}$ of a few
hundred K.  The observed objects galaxies were selected to have a young stellar
population component.  Such a young stellar population might indicate the
presence of significant amounts of gas in the galaxy. A further important
characteristic of the galaxies with fast gaseous outflows is that the radio
activity started (or re-started) very recently in these galaxies ($\lta 10^6$
years ago).

Gas outflows are also detected in quasars and Seyfert galaxies.  For quasars,
Elvis (2000) suggested they are connected with regions close to the
accretion-disk. In Seyferts, radiation and/or wind pressure may drive
the outflows (Dopita et al.\ 2002) although in radio-loud Seyferts they
could be due to jet-cloud interactions (e.g.\ Oosterloo et al.\ 2000). 
Starburst winds are  detected in a number of nearby galaxies (Heckman et
al.\ 1990).

\begin{figure}
\centerline{\psfig{figure=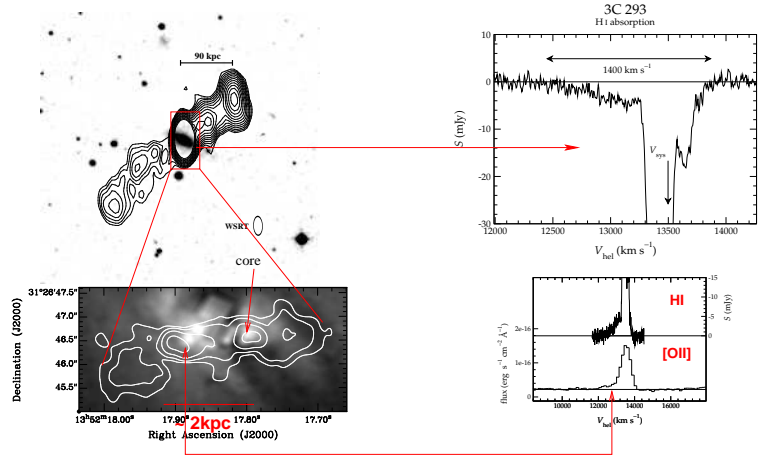,angle=0,width=10cm}
}
\caption{The data for the nearby radio galaxy 3C~293. {\sl Top left:} the large scale 
radio continuum.  {\sl Bottom left:} High resolution radio contours of the
central region (MERLIN observations from Beswick et al.\ 2002) superimposed to
an HST image. {\sl Top right:} zoom-in of the WSRT \HI\ absorption profile,
where the broad, shallow absorption is clearly visible (Morganti et al.\
2003). {\sl Bottom right:} comparison between the \HI\ absorption and the
[\OII]3727\AA\ profile. The similarity of the broad, blueshifted wing in the
two profiles is evident. The broad wing in the optical spectrum is detected at
the position of the E radio lobe, not near the core of the radio galaxy.}
\end{figure}

\section{The case of the nearby radio galaxy 3C~293}

A key observation to discriminate between the mechanisms responsible for gas
outflows  is to identify the location where the outflow (both
in ionised and neutral gas) is occurring.  For 3C~293, new optical spectra
(Emonts et al.\ in prep.) show that also the optical emission lines contain a
broad blueshifted component that is very similar in width to the broad \HI\
absorption. This optical outflow {\sl is detected at the position of the
eastern radio lobe, 1 kpc from the nucleus} (see Fig.\ 1). The
similarity in the spectra suggests that the \HI\ and ionised gas outflows have
a common origin. The fact that the broad optical outflow is not detected near
the nucleus but near a radio lobe, may imply that the interaction between the
radio plasma and the ISM is responsible for the outflow.  A possible scenario
is that the radio jet hits molecular or neutral clouds in the ISM.  These
clouds fragment and are accelerated and ionised by the jet shocks.  Due to
efficient cooling, dense fragments will quickly recombine to form dense, fast
moving clouds of neutral (and molecular) gas.  Mellema et al.\ (2002) proposed
such a mechanism to explain the phenomenon of jet-induced star formation in



\begin{thebibliography}

  \bibitem[]{}
     {Beswick A.J., Pedlar A., Holloway A.J.} 2002,
     \textit{MNRAS}, \textbf{369}, 620
\bibitem[]{}{Dopita M.A. et al.}, 2002, \textit{ApJ}, \textbf{572}, 753
\bibitem[]{}{Elvis M.} 2000, \textit{ApJ}, \textbf{545}, 63
\bibitem[]{} {Evans A.S., Sanders D.B., Surace J.A., Mazzarella J.M.} 1999,
\textit{ApJ}. \textbf{511}, 730 
\bibitem[]{}{Heckman T.M., Armus
L., Miley G.K.} 1990, \textit{ApJS}, \textbf{74}, 833
 \bibitem[]{}{Mellema, G.  Kurk J.D., R\"ottgering H.}  2002, \textit{A\&A}, \textbf{395}, L1
 \bibitem[]{}{Morganti R., Oosterloo T., Emonts B., van der Hulst J.M.,
Tadhunter C.} 2003, \textit{ApJ}, \textbf{593}, L69
\bibitem[]{}{Oosterloo T. et al.} 2000, \textit{AJ}, \textbf{119}, 2085
\end{thebibliography}
\end{document}